\documentclass[12pt, eqno]{article}
\usepackage[dvips]{color}
\usepackage{epsfig}
 \usepackage{amsmath}
\usepackage{graphicx}
\def\Box{\hbox{$\rlap{$\sqcup$}\sqcap$}}
\begin{document}
\numberwithin{equation} {section}
\title{\bf Bouncing Universe and Reconstructing Vector Field }
\author{{J. Sadeghi $^{a}$\thanks{Email:
pouriya@ipm.ir}\hspace{1mm}, M. R. Setare $^{b}$ \thanks{Email:
rezakord@ipm.ir}\hspace{1mm}, A. R. Amani$^{c}$ \thanks{Email:
a.r.amani@iauamol.ac.ir}}\ and S. M. Noorbakhsh$^{a}$\thanks{Email:
s.m.noorbakhsh@umz.ac.ir} \\ $^a${\small {\em Sciences Faculty,
Department of Physics, Mazandaran University,}}\\{\small {\em P .O
.Box 47415-416, Babolsar, Iran}}\\$^b${\small {\em Department of Science, Payame Noor University, Bijar, Iran}}\\
$^c$ {\small {\em  Department of Physics, Islamic Azad University - Ayatollah Amoli Branch,}}\\
        {\small {\em P.O.Box 678, Amol, Iran}}}
\maketitle

\begin{abstract}
Motivated by the recent works of Refs. \cite{R1, R2} where a model
of inflation has been suggested with non-minimally coupled massive
vector fields, we generalize their work to the  study of the
bouncing solution. So we consider a massive vector field, which is
non-minimally coupled to gravity. Also we consider non-minimal
coupling of vector field to the scalar curvature. Then we
reconstruct this model in the light of three forms of
parametrization for dynamical dark energy. Finally we simply
plot reconstructed physical quantities in flat universe.\\

{\bf Keywords:} Massive vector field; Bouncing; Reconstruction;
Parametrization.\\

{\bf PACS:} 98.80.Cq, 98.80.-k, 98.80.Jk.
\end{abstract}
\section{Introduction}
Nowadays it is strongly believed that the universe is experiencing
an accelerated expansion. The observation data  confirm it such
as type Ia supernovae \cite{R3} in associated with large scale
structure \cite{R4} and Cosmic Microwave Background anisotropies
\cite{R5} have provided main evidence for this cosmic
acceleration. In order to explain why the cosmic acceleration
happens, many theories have been proposed. The standard
cosmological model (SCM) furnishes an accurate description of the
evolution of the universe, in spite of its success, the SCM
suffers from a series of problems such as the initial
singularity, the cosmological horizon, the flatness problem, the
baryon asymmetry and the nature of dark energy and dark matter,
although inflation partially or totally answers some of these
problems. Inflation theory was first proposed by Guth in 1981
\cite{gu}. Inflation is a period of accelerated expansion in the
early universe, it occurs when the energy density of the universe
is dominated by the potential energy of some scalar field called
inflaton. Currently all successful inflationary scenarios are
based on the use of weakly interaction scalar fields. Scalar
fields naturally arise in particle physics including string theory
and these can act as candidates for dark energy. So far a wide
verity of scalar field dark energy models have been proposed.
These include quintessence \cite{R6}, K-essence \cite{R7},
 tachyon \cite{R8}, phantoms \cite{R9}, ghost condensates
\cite{R10} and so forth. Two main reason for use of scalar fields
to explain inflation are natural homogeneity and isotropy of such
fields and its ability to imitate a slowly decaying cosmological
constant \cite{R1}. However, no scalar field has ever been
observed, and designing models by using unobserved scalar fields
undermine their predictability and falsifiability, despite the
recent precision data. The latest theoretical developments (string
landscape) offer too much freedom for model-building, so higher
spin fields generically induce a spatial anisotropy and the
effective mass of such fields usually of the order of the Hubble
scale and the slow-roll inflation dose not occurs \cite{R11}.
Then an immediate question is, can we do Cosmology without scalar
fields? The authors of \cite{R1,R2} have shown that a successful
vector inflation can be simultaneously surmounted in a natural
way, and isotropy of the vector field condensate be achieved
either in the case of triplet of mutually orthogonal vector field
\cite{R12}. In spite of inflation success in explaining the
present state of the universe, it dose not solve the crucial
problem of the initial singularity \cite{R13}. The existence of
an initial singularity is disturbing, because the space-time
description breaks down "there". Non-singular universes have been
recurrently present in the scientific literature. Bouncing model
is one of them that was first proposed by Novello and Salim
\cite{R14} and Mlnikov and Orlov \cite{R15} in the late 70's. At
the end of the 90's the discovery of the acceleration of the
universe brought back to the front the idea that $\rho+3p$ could
be negative, which is precisely one of the conditions needed for
cosmological bounce in GR, and contributed to the revival of
nonsingular universes. Bouncing universe are those that go from
an era of acceleration collapse to an expanding era without
displaying a singularity \cite{R16}. Necessary conditions
required for a successful bounce during the contracting phase,
the scale factor $a(t)$ is decreasing , i.e. $\dot{a} < 0$, and
in the expanding phase we have $\dot{a} > 0$. At the bouncing
point, $\dot{a} = 0$, and around this point $\ddot{a} > 0$ for a
period of time. Equivalently in the bouncing cosmology the Hubble
parameter H runs across zero from $\dot{H} < 0$ to ${H}> 0$ and
$H = 0$ at the bouncing point. A successful bounce
requires around this point.\\
The remainder of the paper is as follows. In section 2 and 3, we
will consider vector field action where proposed in Refs.
\cite{R1,R2} and study bouncing solution of this model. In
section 4 and 5,
 we will reconstruct physical quantities for this model and also will plot the corresponding graphs. Finally we will
 apply three parametrization and compare them for this model.\\

\section{Vector field foundation}
We consider a massive vector field, which is non-minimally coupled
to gravity, \cite{R1,R2}. The action is given by

\begin{equation}\label{E1}
S=\int d^{4}x \sqrt{-g}~\left(\frac{1}{16 \pi G}R-\frac{1}{4}
F_{\mu\nu}F^{\mu \nu}-\frac{1}{2}m^2 U_{\mu}U^{\mu}+\frac{1}{2}\xi R
U_{\mu}U^{\mu}\right),
\end{equation}
where $F_{\mu\nu}=\partial_{\mu}U_{\nu}-\partial_{\nu}U_{\mu}$, and
$\xi$ is a dimensionless parameter for non-minimal coupling. We note
that, the non-minimal coupling of vector field is same with
conformal coupling of a scalar field in case $\xi=1/6$. We adopt FRW
universe with the metric signature of $(-+++)$.\\
The equations of motion are given by
\begin{eqnarray}\label{E2}
R_{\mu\nu}-\frac{1}{2}R g_{\mu\nu}={8 \pi
G}\bigg[F_{\mu\alpha}F^{\alpha\nu}-\frac{1}{4}g_{\mu\nu}F_{\alpha\beta}F^{\alpha\beta}+\
(m^2 - \xi R )U_{\mu}U_{\nu}\nonumber\\-\frac{1}{2}g_{\mu\nu}(m^2 -
\xi R )U_{\alpha}U^{\alpha}\ - \xi g_{\mu\nu}U_{\alpha}U^{\alpha} +
\xi (\nabla_{\mu}\nabla_{\nu}-g_{\mu\nu}\Box)U_{\alpha}U^{\alpha}
\bigg],
\end{eqnarray}

\begin{equation}\label{E3}
\nabla_{\nu}F^{\nu\mu} - m^2U^{\mu} + \xi R U^{\mu}=0.
\end{equation}
Where the right hand side of equation (\ref{E2}) is the
energy-momentum tensor of the vector field $U_i$. The variation
of the action with respect to $U_i$ yields the following
equations of motion,
\begin{equation}\label{E4}
\frac{1}{a^2}\nabla ^2 U_0 - \frac{1}{a^2}\partial_i
\dot{U_i}-m^2U_0 +\xi R U_0 =0,
\end{equation}
\begin{equation}\label{E5}
\ddot{U_i}+\frac{\dot{a}}{a}(\dot{U_i}-
\partial_iU_0)-\partial_i\dot{U_i}+\frac{1}{a^2}(\partial_i(\partial_kU_k)-\nabla^2 U_i)
+ m^2U_i -\xi RU_i=0,
\end{equation}
Where $a$ is the scale factor, the dot denotes the derivative with
respect to the cosmic time and  the summation over repeated spatial
indices is satisfied. By considering the quasi-homogeneous vector
field $(\partial _iU_\alpha = 0)$ and Eq. (\ref{E4}) imply $U_0 =
0$, so that from Eq. (\ref{E5}) we obtain
\begin{equation}\label{E6}
\ddot{U_i}+ H\dot{U_i}-6\xi(\dot{H}+2H^2+\frac{k}{a^2})U_i+m^2U_i=0.
\end{equation}
By using acceleration relation $\frac{\ddot{a}}{a}=-\frac{4\pi
G}{3}(\rho +3p)$ we achieve as,
\begin{equation}\label{E7}
\dot{H}+H^2=\frac{-4\pi G}{a^2}\left(2\dot{U_i^2}-4(1+6\xi)H
U_i\dot{U_i}+6\xi U_i^2H^2 -m^2U_i^2 -\frac{2k}{a^2}\xi
U_i^2\right),
\end{equation}
where $H=\frac{\dot{a}}{a}$, $R = 6(\dot{H}^2+2H^2+\frac{k}{a})$
and $R^0_0=\dot{H}+H^2$ are Hubble 's parameter, Ricci scalar and
first component of Ricci tensor, respectively. As we know, a
dynamical vector field has generally a preferred direction, and
to introduce such a vector field may not be consistent with the
isotropy of the universe. In fact, the energy-momentum tensor of
the vector field $U_\mu$  has anisotropic components. However,
the anisotropic part of the energy-momentum tensor can be
eliminated by introducing a triplet of mutually orthogonal vector
fields. In that case, we obtain the energy density $\rho$ and the
pressure $p$ of the vector fields
\begin{equation}\label{E8}
\rho=\frac{1}{a^2}\left[\frac{3}{2}\dot{U_i}^2-3(1+6\xi)HU_i\dot{U_i}+9\xi
U_i^2H^2 +\frac{3}{2}m^2U_i^2-\frac{9k\xi}{a^2}U_i^2\right],
\end{equation}
\begin{equation}\label{E9}
p=\frac{1}{a^2}\left[\frac{3}{2}\dot{U_i}^2-3(1+6\xi)HU_i\dot{U_i}+9\xi
U_i^2H^2 -\frac{3}{2}m^2U_i^2+\frac{3k\xi}{a^2}U_i^2\right].
\end{equation}
Now to introduce a change of variable $\phi_i=\frac{U_i}{a}$ (for
more detail see Ref. [1]), equation (\ref{E4}) make change to,
\begin{equation}\label{E10}
\ddot{\phi_i}+3H\dot{\phi}+\left(m^2+(1-6\xi)(\dot{H}+2H^2)-\frac{6\xi
k}{a^2}\right)\phi_i=0.
\end{equation}
Then we consider  $\xi=1/6$ and obtain the basic equations of motion
for a curved universe in terms of $\phi_i$,
\begin{equation}\label{E11}
\ddot{\phi_i}+3H\dot{\phi}+\left(m^2-\frac{k}{a^2}\right)\phi_i=0,
\end{equation}
\begin{equation}\label{E12}
H^2+\frac{k}{a^2}=4\pi G(\dot{\phi^2_i}+m^2 \phi^2_i
-\frac{k}{a^2}\phi^2 _i ),
\end{equation}
\begin{equation}\label{E13}
\dot{H}+H^2=-4\pi G(2\dot{\phi^2_i}-m^2 \phi^2_i),
\end{equation}
One can see where equations of motion of vector field is reduced to
minimally coupled massive scalar fields. So energy density $\rho$
and the pressure $p$ for the vector fields are derived in terms of
$\phi_i$ in case $\xi=1/6$ in the form,
\begin{equation}\label{E14}
\rho=\frac{3}{2}\dot{\phi_i}^2+\frac{3}{2}m^2 \phi^2_i
-\frac{3k}{2a^2}\phi^2_i,
\end{equation}
\begin{equation}\label{E15}
p=\frac{3}{2}\dot{\phi_i^2}-\frac{3}{2}m^2 \phi^2_i
+\frac{k}{2a^2}\phi^2_i,
\end{equation}
\\
Now we are going to consider behavior of the different values of
parameter $\xi$ for vector field. We solve numerically Eq.
(\ref{E6}) for $K=0, +1, -1$ which implies the flat, close and open
universe respectively. The Fig.1 shows graph of the vector field
with respect to time in all of cases $K$. One can see where vector
field has oscillation behavior and the magnitude slowly decrease
with respect to time evolution. Also by increasing the parameter
$\xi$, the magnitude of vector field will increase, but the period
of oscillation is constant. We note that negative values of $\xi$
actually is the same of above result.
\begin{figure}[th]
\centerline{\epsfig{file=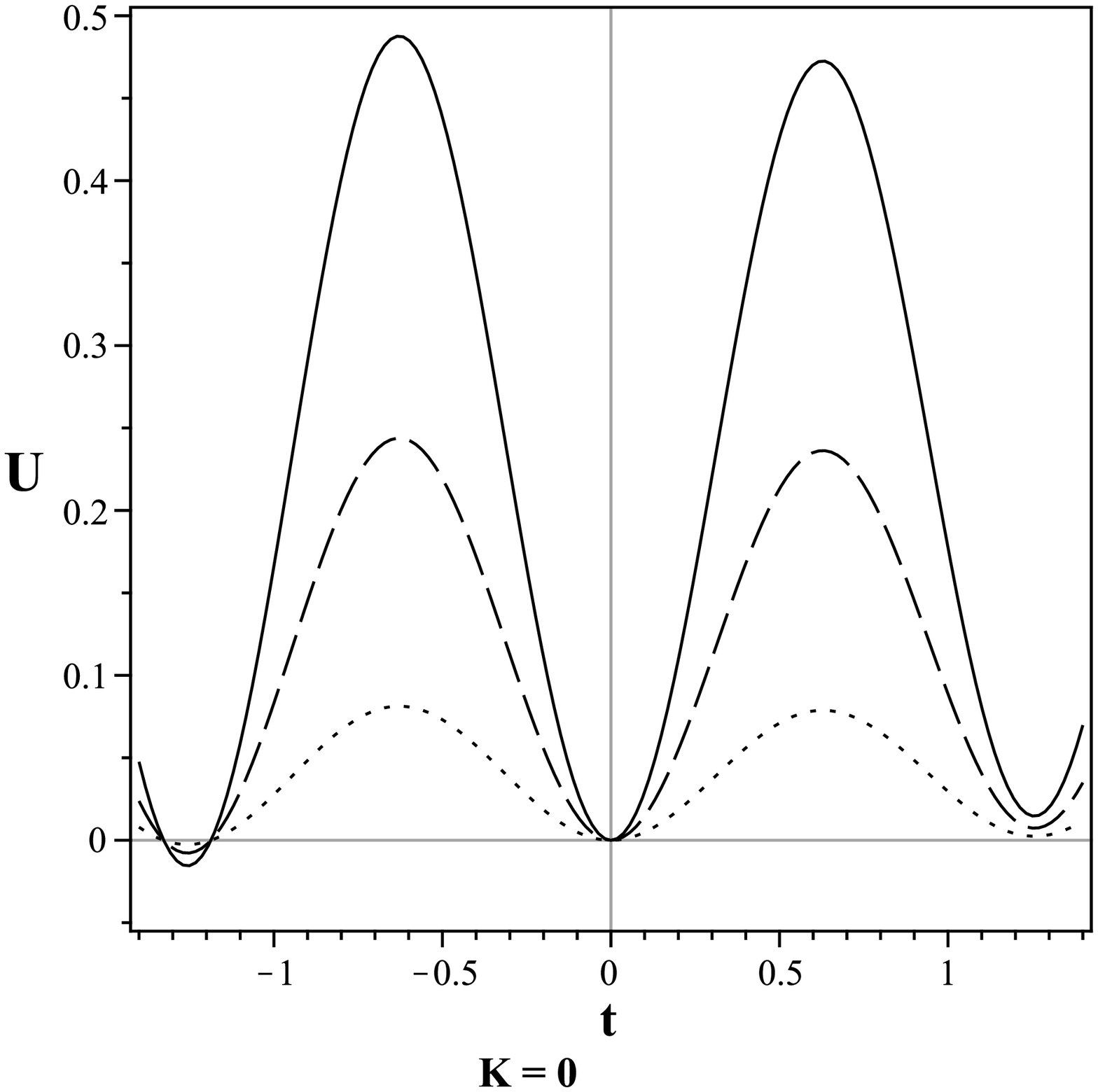,scale=.25}\epsfig{file=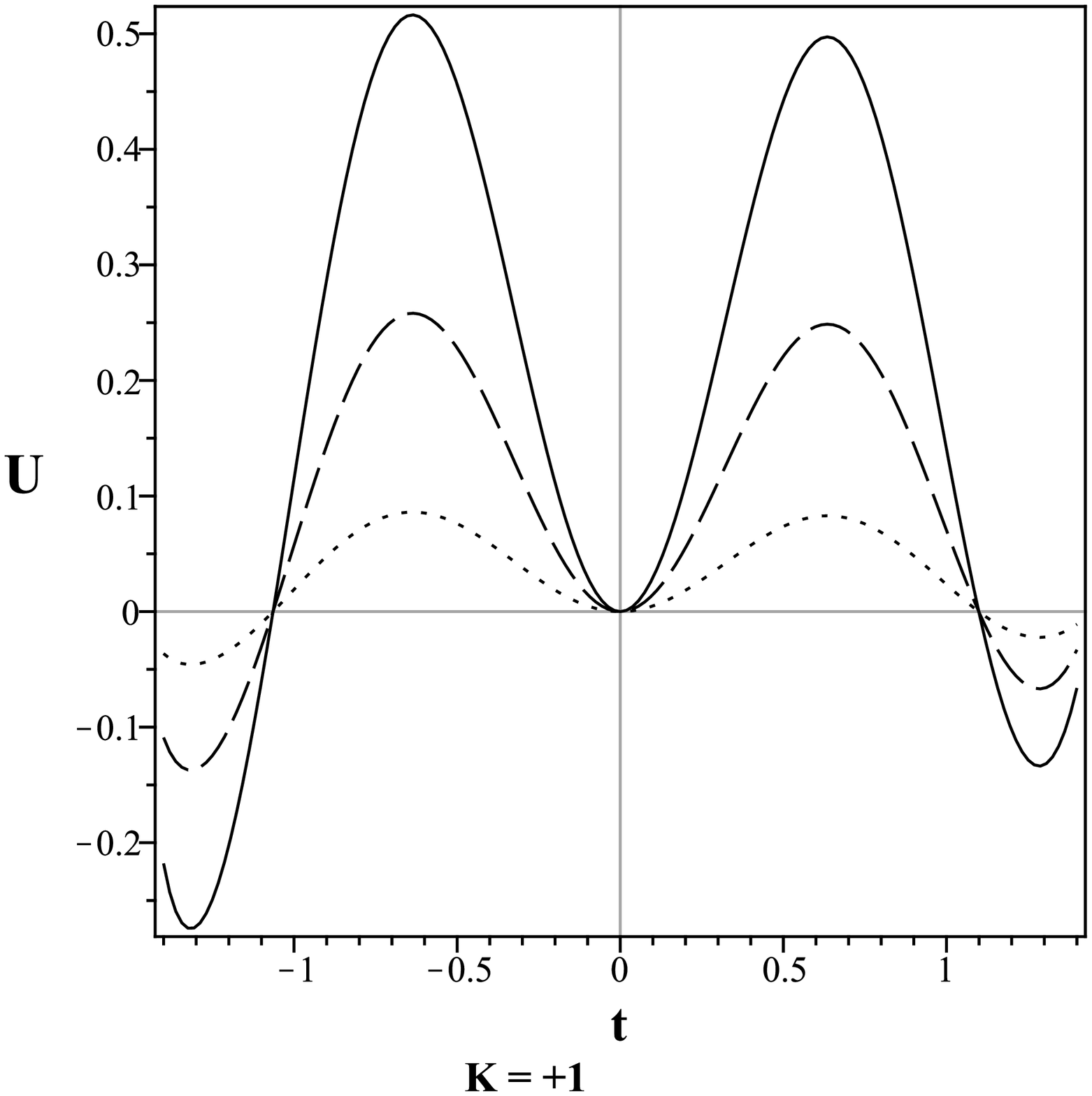,scale=.25}\epsfig{file=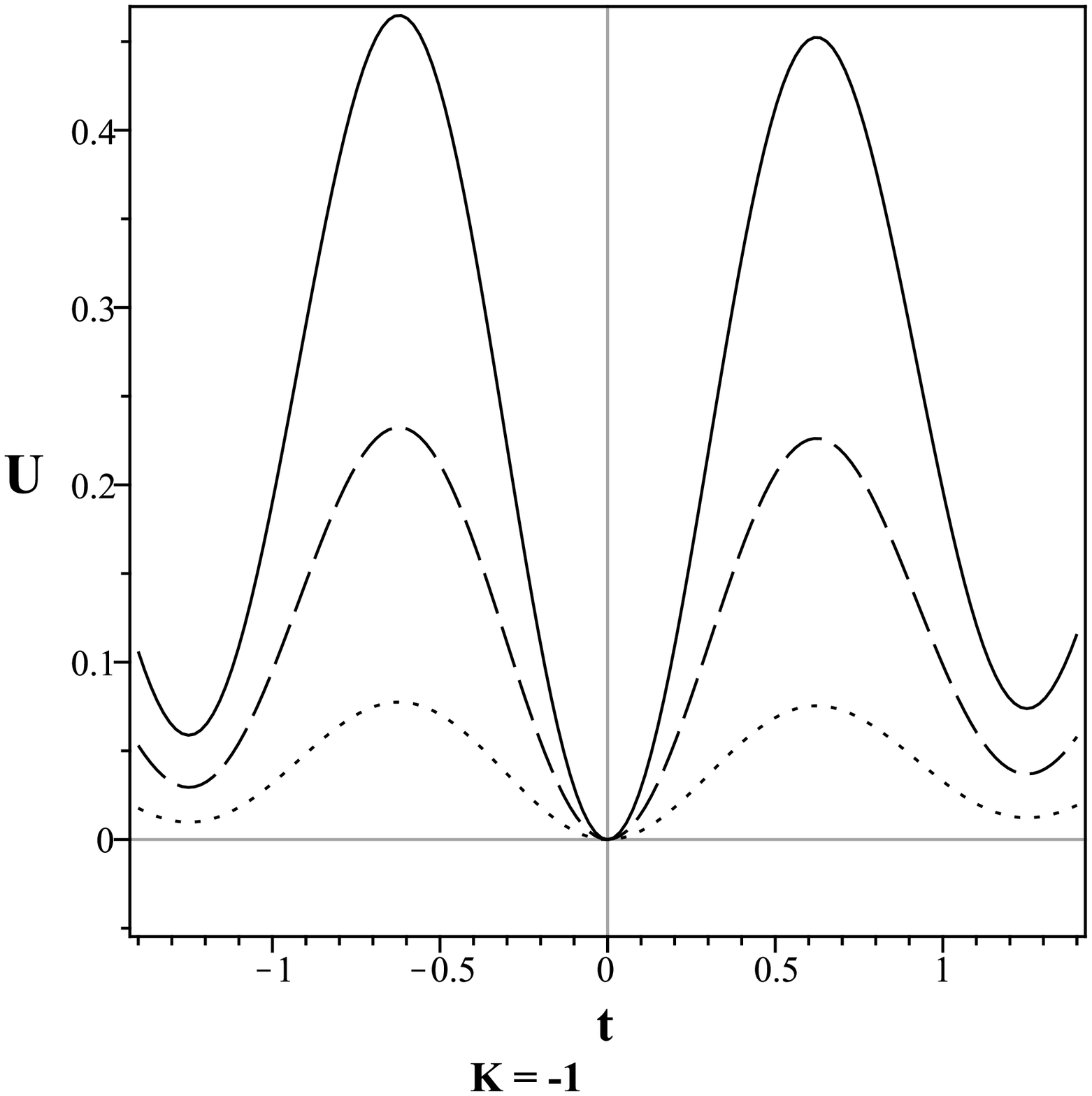,scale=.25}}
\vspace*{8pt} \caption{Graphs of vector fields in term of time. The
solid, dash and doted lines represent $\xi=1$ ,$\frac{1}{6}$ and
$0.5$ respectively.}
\end{figure}

As above mention we suggest following solution for $U_i(t)$
\begin{equation}\label{E20}
U_i(t)=\sqrt{A}e^{-\gamma t}\cos (mt+\theta),
\end{equation}
where the parameter $A$ describes the oscillating amplitude of the
field with dimension of $[mass]^2$. Also $A$ is relation with the
parameter $\xi$, this solution implies the damping magnitude of the
oscillating vector field.

\section{Bouncing behavior}
We will start with a detailed examination on the necessary
conditions required for a successful bounce. During the
contracting phase, the scale factor $a(t)$ is decreasing, i.e.,
$\dot{a}<0$, and in the expanding phase we have $\dot{a}>0$. At
the bouncing point $\dot{a}=0$ and around this point $\ddot{a}>0$
for a period of time. Equivalently in the bouncing cosmology the
Hubble parameter $H$ runs across zero from $H<0$ to $H>0$ and $H =
0$ at the bouncing point. A successful bounce requires around
this point
\begin{equation}\label{E16}
\dot{H}=-4\pi G(\rho + p)+\frac{k}{a^2} > 0.
\end{equation}
At the point where the bounce occurs, Eqs. (\ref{E8}) and (\ref{E9})
reduce to
\begin{equation}\label{E17}
\rho_b=\frac{3}{2a^2}(\dot{U_i}^{2}+m^2U^2_i)-\frac{9k}{a^4}\xi
U_i^2,
\end{equation}
\begin{equation}\label{E18}
p_b=\frac{3}{2a^2}(\dot{U_i}^{2}-m^2U^2_i)+\frac{3k}{a^4}\xi
U_i^2,
\end{equation}
On the other hand, a successful bounce from Eqs. (\ref{E6}),
(\ref{E7}) and (\ref{E16}) obtain in the form,
\begin{equation}\label{E19}
\dot{U_i}^2 < \frac{1}{2}m^2U^2_i+\frac{k}{a^2}\xi U_i^2.
\end{equation}

This result is similar to slow roll inflation. This means that one
requires a flat potential where give rise to a point bounce for
the model of vector field. From conditions (\ref{E16}),
(\ref{E19}) it is clear that if we have bouncing solutions in
open universe, then we have such behaviour for flat and closed
universe as well.
 Now we solve above equation
numerically by different value of $\xi$ on the curved universe
that is plotted in Fig. 2.\\\\
\begin{figure}[th]
\centerline{\epsfig{file=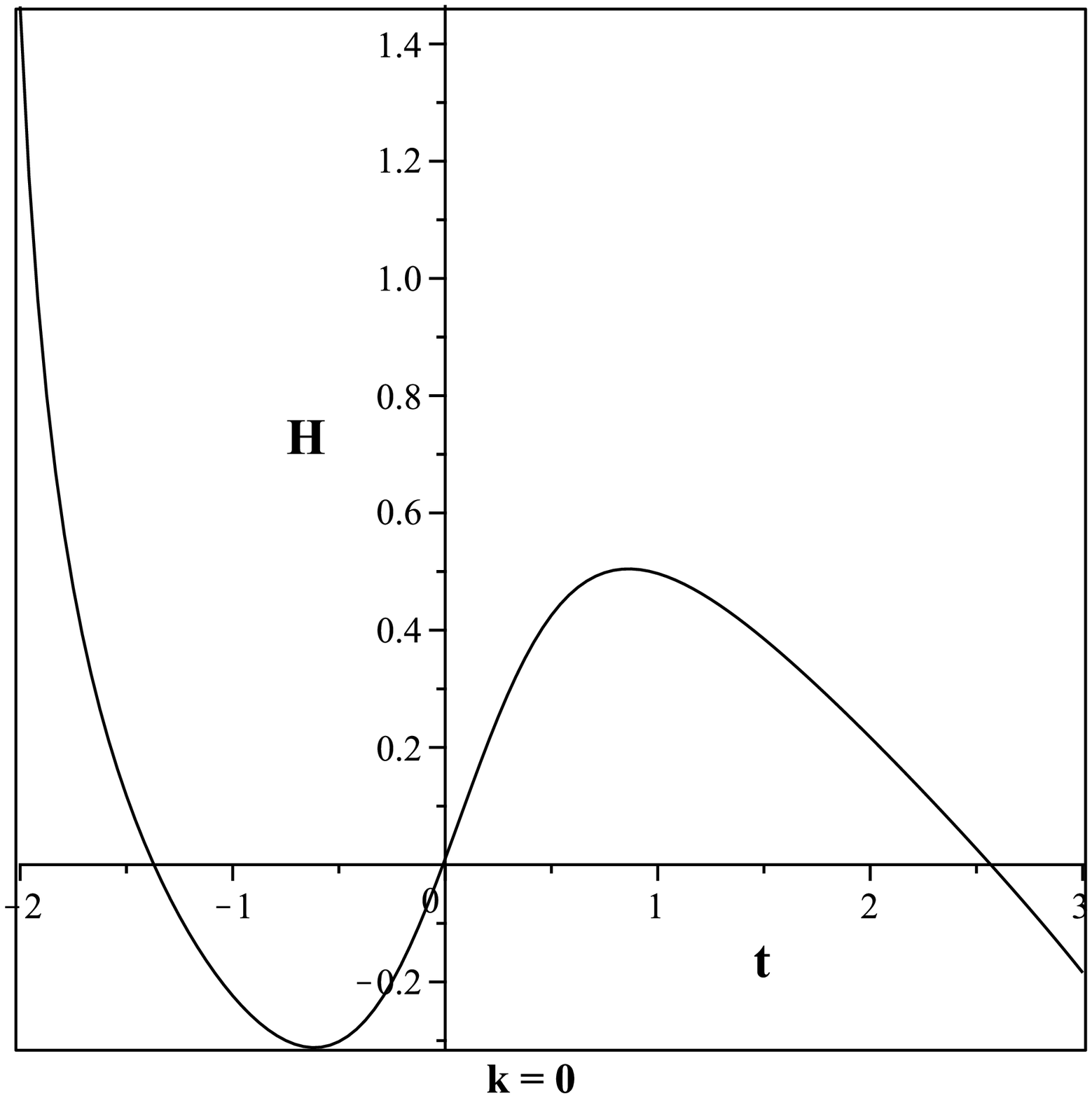,scale=.25}\epsfig{file=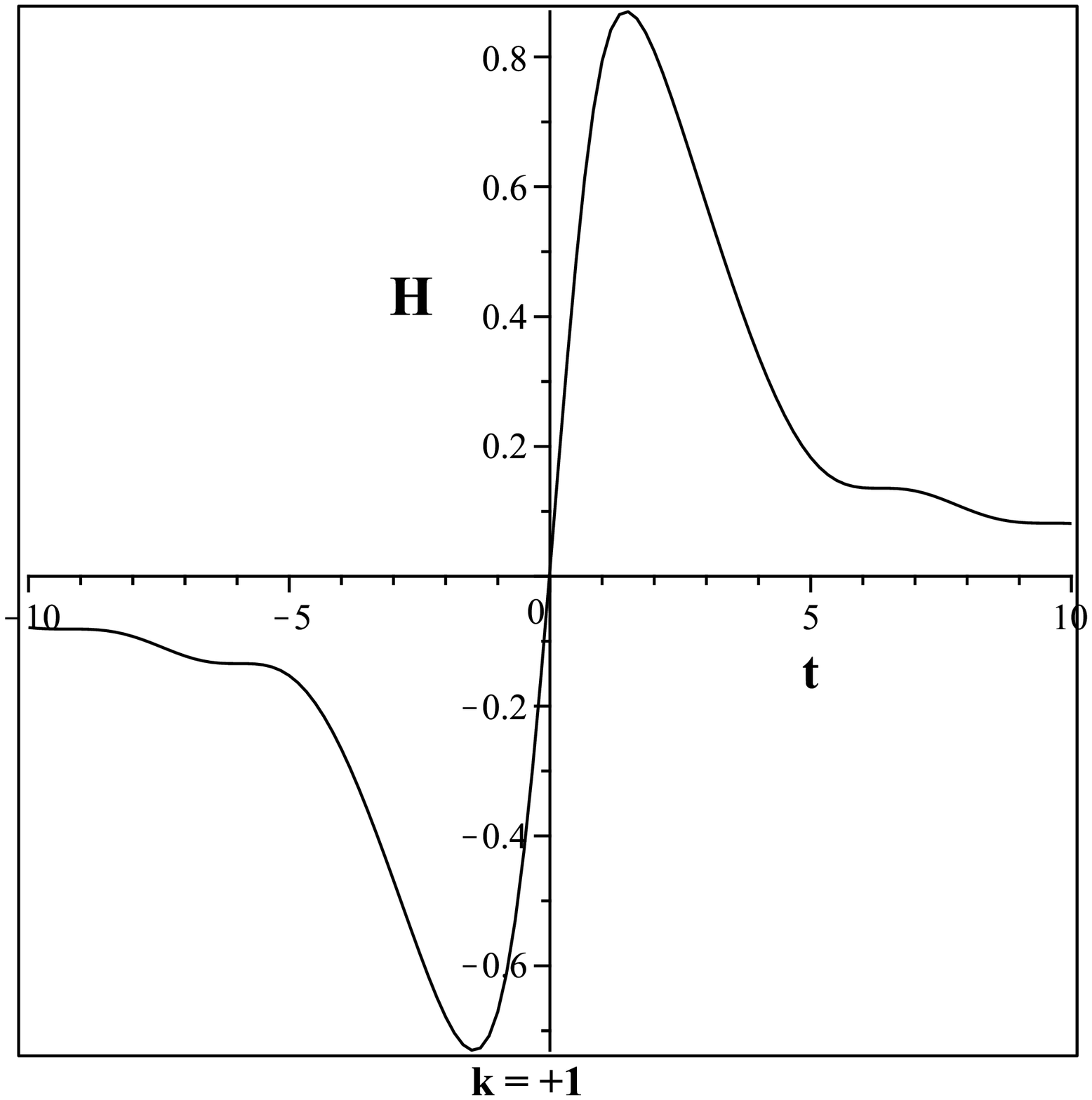,scale=.25}\epsfig{file=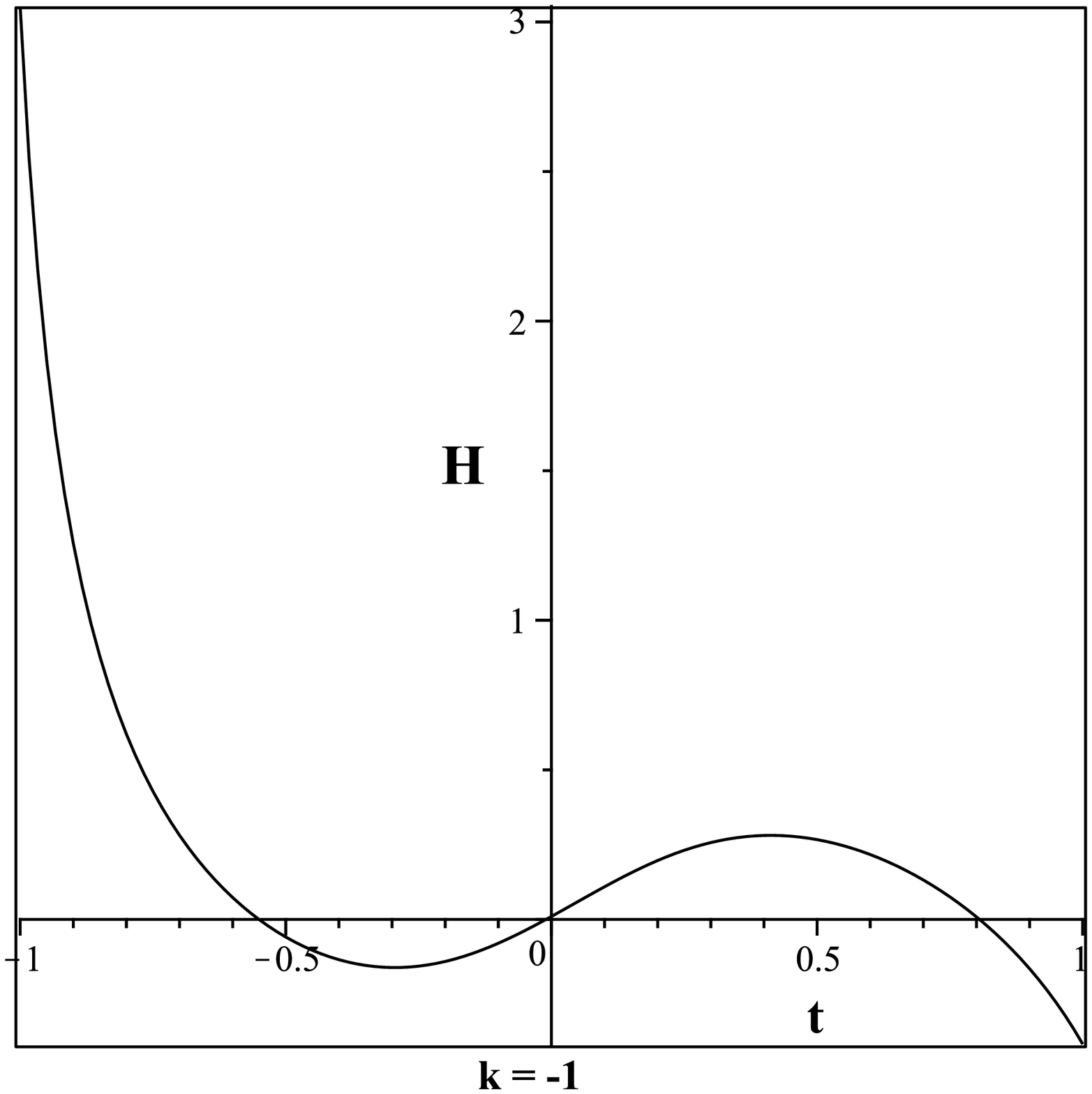,scale=.25}}
\vspace*{8pt} \caption{The graphs of the Hubble's parameter for $\xi=1/6$, $4\pi G=1$, $m=1$ and $k=0,+1,-1$ by choosing $\phi(0)=1$, $\dot{\phi}(0)=0.1$, $a(0)=1$ and $H(0)=0.01$.}
\end{figure}

\begin{figure}[th]
\centerline{\epsfig{file=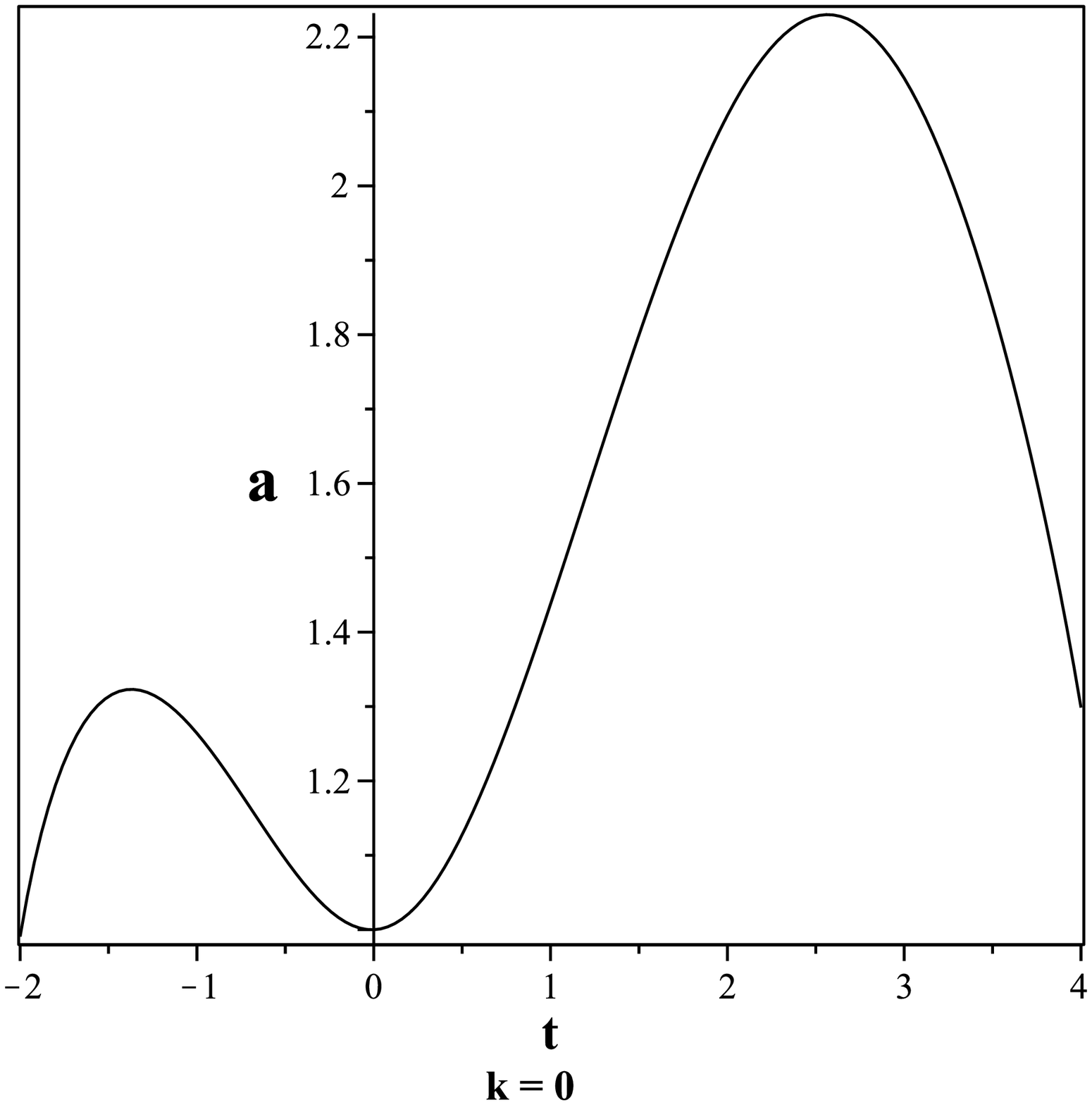,scale=.25}\epsfig{file=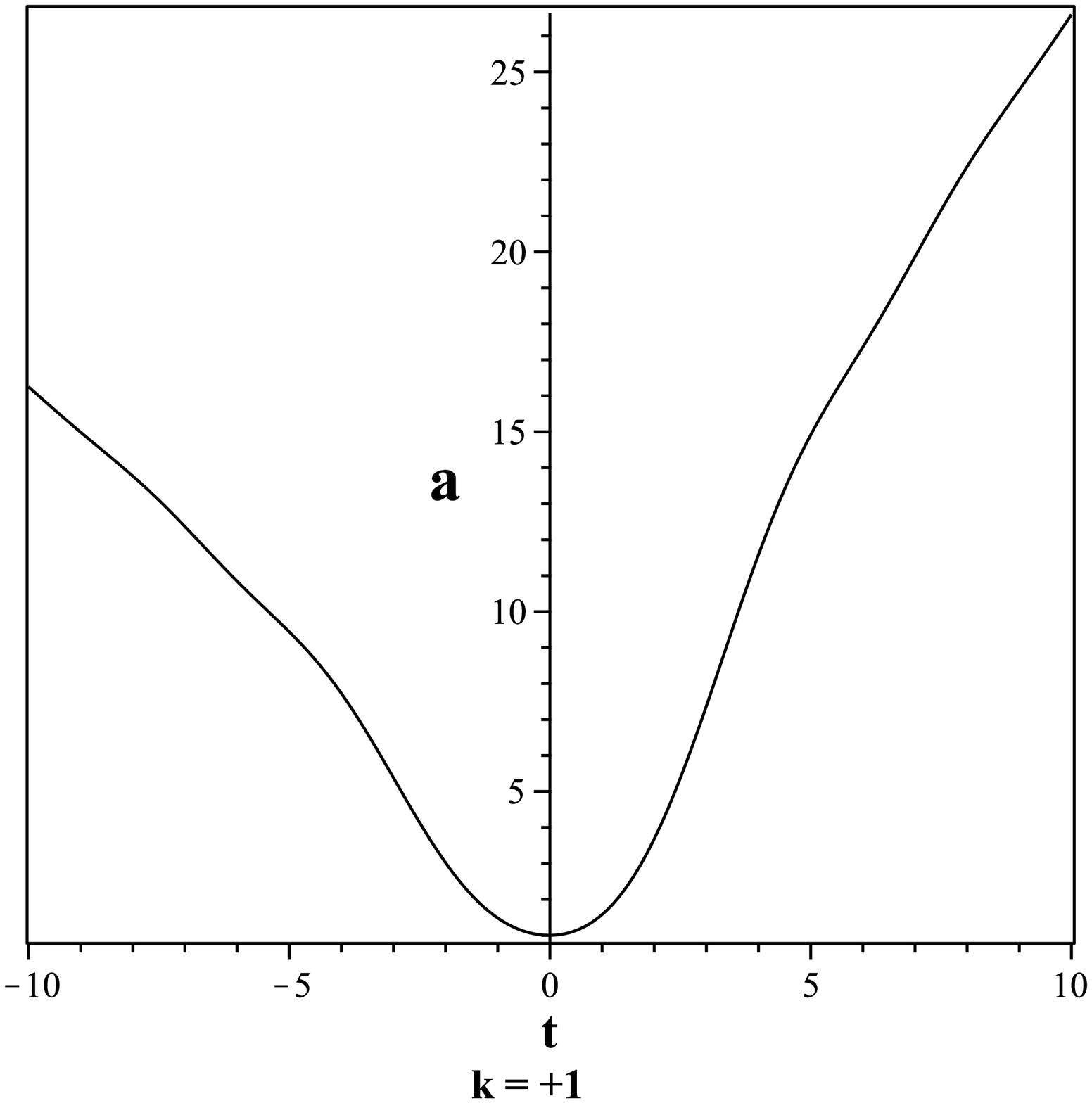,scale=.25}\epsfig{file=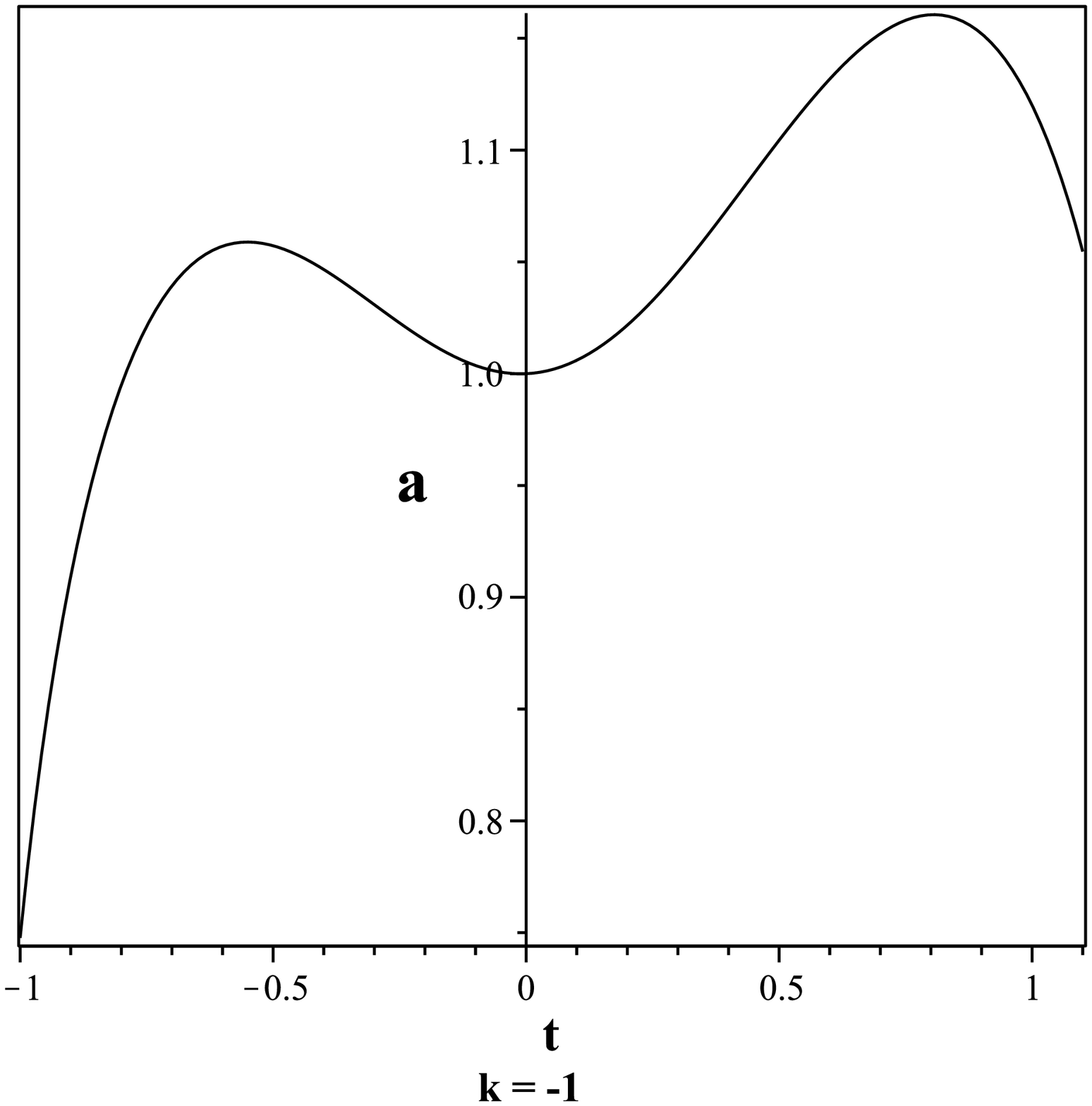,scale=.25}}
\vspace*{8pt} \caption{The graphs of the scale factor for $\xi=1/6$, $4\pi G=1$, $m=1$ and $k=0,+1,-1$ by choosing $\phi(0)=1$, $\dot{\phi}(0)=0.1$, $a(0)=1$ and $H(0)=0.01$.}
\end{figure}

One can see the Hubble parameter $H$ running across zero in any
three cases of $k$. In all cases of  $k$, we have $H<0$ to $H>0$
where implies to go from collapse era to an expanding era, and
this result will not change for the different values $\xi$ in all
of $k$. Also in Fig.3, we can see the behaviour of scale
 factor in terms of time for different values of $k$.
 It is clear that during the contracting phase, the scale factor $a(t)$ is decreasing,
 i.e., $\ddot{a}<0$, and in the expanding phase we have $\ddot{a}>0$, so the point where
$\ddot{a}=0$ is bouncing point. \\
Therefore, in the vector field dominated universe we have a
successful bouncing point in close and flat universe but a
turn-around point in open universe. The bounce can be attributed
to the negative-energy matter, which dominates at small values of
$a$ and create a significant enough repulsive force so that a big
crunch is avoided.
\section{Reconstruction}
Now we are going to present a reconstruction process for vector
field in the curved universe by $\xi=1/6$ . In this section,
potential and kinetic energy are reconstructed with respect to
redshirt $z$. Also we obtain the EoS in term of $z$. After that
three type parametrization are represented for the EoS. By using
it we consider cosmology solutions such as the Eos, the
deceleration parameter and vector field. The stability condition
of this system is described by quantity of the sound speed. We
rewrite Eqs. (\ref{E14}) and (\ref{E15}) in term of the effective
potential energy $\hat{V}$ and the effective kinetic energy
$\hat{K}$ as the following form,
\begin{equation}\label{E21}
\rho=\frac{3}{2}\dot{\phi_i}^2+\frac{3}{2}m^2 \phi^2_i
-\frac{3k}{2a^2}\phi^2_i=3\hat{K}+3\hat{V},
\end{equation}
\begin{equation}\label{E22}
p=\frac{3}{2}\dot{\phi_i}^2-\frac{3}{2}m^2 \phi^2_i
+\frac{k}{2a^2}\phi^2_i=3\hat{K}-3\hat{V}-\frac{k}{a^2},
\end{equation}
\begin{equation}\label{E23}
\rho +p=6\hat{K}-\frac{k}{a^2}.
\end{equation}
Then we can write the Friedmann equations as following
\begin{equation}\label{E24}
3M^2_p (H^2+\frac{k}{a^2})=\rho_m+\rho=\rho_m+3\hat{K}+3\hat{V},
\end{equation}
\begin{equation}\label{E25}
2M^2_p(\dot{H}-\frac{k}{a^2})=-\rho_m-\rho-p=-\rho_m-6\hat{K}+\frac{k}{a^2},
\end{equation}
where $\rho_m$ is the energy density of dust matter. Also from Eqs.
(\ref{E21}) and (\ref{E22}), we obtain relationship between the Eos
with $\hat{V}$ and $\hat{K}$ as,
\begin{equation}\label{E26}
\omega=\frac{p}{\rho}=\frac{3\hat{K}-3\hat{V}-\frac{k}{a^2}}{3\hat{K}+3\hat{V}}=-1+\frac{2-\frac{k}{3a^2\hat{K}}}{1+\frac{\hat{V}}{\hat{K}}}.
\end{equation}
We obviously have
\begin{eqnarray}\label{E26-1}
\hat{V}+3\hat{K}>\frac{k}{3a^2} \Longrightarrow \omega > -1,\nonumber\\
\hat{V}+3\hat{K}<\frac{k}{3a^2} \Longrightarrow \omega < -1,\nonumber\\
\hat{V}+\hat{K}=\frac{k}{3a^2} \Longrightarrow \omega=-1.
\end{eqnarray}
By using Eqs. (\ref{E24}) and (\ref{E25}) we can write
\begin{equation}\label{E27}
\hat{K}=\frac{-\rho_m}{6}-\frac{M^2_p}{3}(\dot{H}-\frac{k}{3a^2})+\frac{k}{6a^2},
\end{equation}
\begin{equation}\label{E28}
\hat{V}=\frac{M^2_p}{3}(3H^2+\dot{H}+\frac{2k}{a^2})-\frac{\rho_m}{6}-\frac{k}{6a^2}.
\end{equation}
As in the present model, the dark energy fluid does not couple to
the background fluid, the expression of the energy density of dust
matter in respect of redshift $z$ is \cite{R17},
\begin{equation}\label{E29}
\rho_m=3M^2_pH^2_0\Omega_{m_0}(1+z)^3,
\end{equation}
where $\Omega_{m_0}$ is the ratio density parameter of matter fluid
and the subscript $0$ indicates the present value of the
corresponding quantity. By using the equation $1+z=\frac{a_0}{a}$
($a_0$ is quantity given at the present epoch) and its differential
form in following have,
\begin{equation}\label{E30}
\frac{d}{dt}=-H(1+z)\frac{d}{dz}.
\end{equation}
To introduce a new variable $r$ as,
\begin{equation}\label{E31}
r=\frac{H^2}{H^2_0},
\end{equation}
we rewrite the equation of motion of vector field against $z$ as,
\begin{eqnarray}\label{E31}
2r(1+z)^2 U^{\prime \prime}_i+2r(1+z)(1+H^2_0)U^{\prime}_i-r^\prime
(1+z)^2U^{\prime}_i\nonumber\\+r^\prime (1+z) U_i-r
U_i+\frac{2m^2}{H^2_0}U_i-\frac{2k}{a^2_0 H^2_0}(1+z)^2 U_i=0,
\end{eqnarray}
$\hat{K}$, $\hat{V}$ can be rewrite as following
\begin{equation}\label{E33}
\hat{K}=-\frac{1}{2}M^2_p H^2_0\Omega_{m0}(1+z)^3+\frac{1}{6}M^2_p
H^2_0r^\prime(1+z)+\frac{k}{6 a_0^2}(1+z)^2,
\end{equation}
\begin{equation}\label{E34}
\hat{V}=M^2_pH^2_0r
+\frac{2k}{3a_0^2}(1+z)^2-\frac{1}{6}M^2_pH^2_0(1+z)r^\prime-\frac{1}{2}M^2_pH^2_0\Omega_{m0}(1+z)^3-\frac{k}{6a_0^2}(1+z)^2.
\end{equation}
By using Eqs. (\ref{E26}), (\ref{E27}) and (\ref{E28}) we obtain
following expression for the EoS,
\begin{equation}\label{E35}
\omega=\frac{(1+z)r^\prime-3r +\frac{k (1+z)^2}{a_0^2 H^2_0
M^2_p}(M^2_p-2) }{3r-3\Omega_{m0}(1+z)^3 +\frac{k (1+z)^2}{a_0^2
H^2_0M^2_p}(M^2_p+2)}.
\end{equation}

Then we obtain following equation for $r(z)$
\begin{eqnarray}\label{E38}
r(z)=\Omega_{m0}(1+z)^3+(1-\Omega_{m0})e^{\beta(z)}\nonumber
\\+\alpha_0 \int^z_0
\left[\omega(\tilde{z})(2+M^2_p)+(2-M^2_p)\right]
(1+\tilde{z})e^{-\beta(\tilde{z})}d\tilde{z}.
\end{eqnarray}
where $\beta(z)=\int^z_0
\frac{3w(\tilde{z})}{1+\tilde{z}}d\tilde{z}$ and
$\alpha_0=\frac{k}{a_0^2 H^2_0 M^2_p}$.
\begin{equation}\label{E38}
r(z)=\Omega_{m0}(1+z)^3+(1-\Omega_{m0})e^{3\int^z_0}\frac{1+w(\tilde{z})}{1+\tilde{z}}d\tilde{z}.
\end{equation}
Also we have following expression for deceleration parameter q
\begin{equation}\label{E39}
q(z)=-1-\frac{\dot{H}}{H^2}=\frac{(1+z)r^\prime -2r}{2r}.
\end{equation}
Now we consider the stability of this model by use the
hydrodynamic analogy and judge on stability by examining  the
value of the sound speed. Of course this is a simple approach,
the perturbations in vector inflation are much richer than in
hydrodynamic model, see recent interesting works in
\cite{{pel1},{pel2}}. The sound speed can be obtained by the
following equation,
\begin{equation}\label{E36}
c^2_s=\frac{p^\prime}{\rho^\prime}=\frac{-2r^\prime+(1+z)r^{\prime\prime}+2\frac{k
(1+z)}{a_0^2 H^2_0 M^2_p}(M^2_p-2)
}{-9\Omega_{m0}(1+z)^2+3r^\prime+2\frac{k(1+z)}{a_0^2 H^2_0
M^2_p}(M^2_p+2)},
\end{equation}
in order to deal the stability of our model, the sound speed must
become $c^2_s\geq0$, so we can obtain from above equation following
condition
\begin{equation}\label{E37}
r(z)\geq \omega_{m0}(1+z)^3 -\frac{16 k a^2_0}{H^2_0(1+z)^2}.
\end{equation}
\section{Parametrization}
Now we consider the three different forms of parametrization as
following and compare them together.\\
 \textbf{Parametrization 1:} First Parametrization has
proposed by Chevallier and Polarski \cite{R18},and Linder
\cite{R19}, where the EoS of dark energy in term of redshift $z$ is
given by,
\begin{equation}\label{E40}
\omega(z)=\omega_0 +\frac{\omega_a z}{1+z}.
\end{equation}
\textbf{Parametrization 2:} Another the EoS in term of redshift z
has proposed by Jassal, Bagla and Padmanabhan \cite{R20} as,
\begin{equation}\label{E41}
\omega(z)=\omega_0 +\frac{\omega_b z}{(1+z)^2}.
\end{equation}
\textbf{Parametrization 3:} Third parametrization has proposed by
Alam, Sahni and Starobinsky \cite{R21}. They take expression of r in
term of $z$ as followoing,
\begin{equation}\label{E41}
r(z)=\Omega_{m0}(1+z)^3+A_0+A_1(1+z)+A_2(1+z)^2.
\end{equation}
By using the results of Refs. \cite{R22}, \cite{R23}, \cite{R24} and
\cite{R25}, we get coefficients of parametrization $1$ as
$\Omega_{m0}=0.29$, $\omega_0= -1.07$ and $\omega_a=0.85$,
 coefficients of parametrization $2$ as $\Omega_{m0}=0.28$,
 $\omega_0=-1.37$ and $\omega_b =3.39$ and coefficients of
parametrization $3$ as $\Omega_{m0}=0.30$, $A_0 = 1$, $A1 = -0.48$
and $A2 = 0.25$. The evolution of $\omega(z)$ and $q(z)$ are plotted
in Fig. 3. Also, using Eqs. (\ref{E33}) and (\ref{E34}) and the
three parametrization, the evolutions of $\hat{K(z)}$ and
$\hat{V(z)}$ are shown in Fig. 5 and Fig. 6 respectively. We note
that graphs simply represent only in the flat universe.
\begin{figure}[th]
\centerline{\epsfig{file=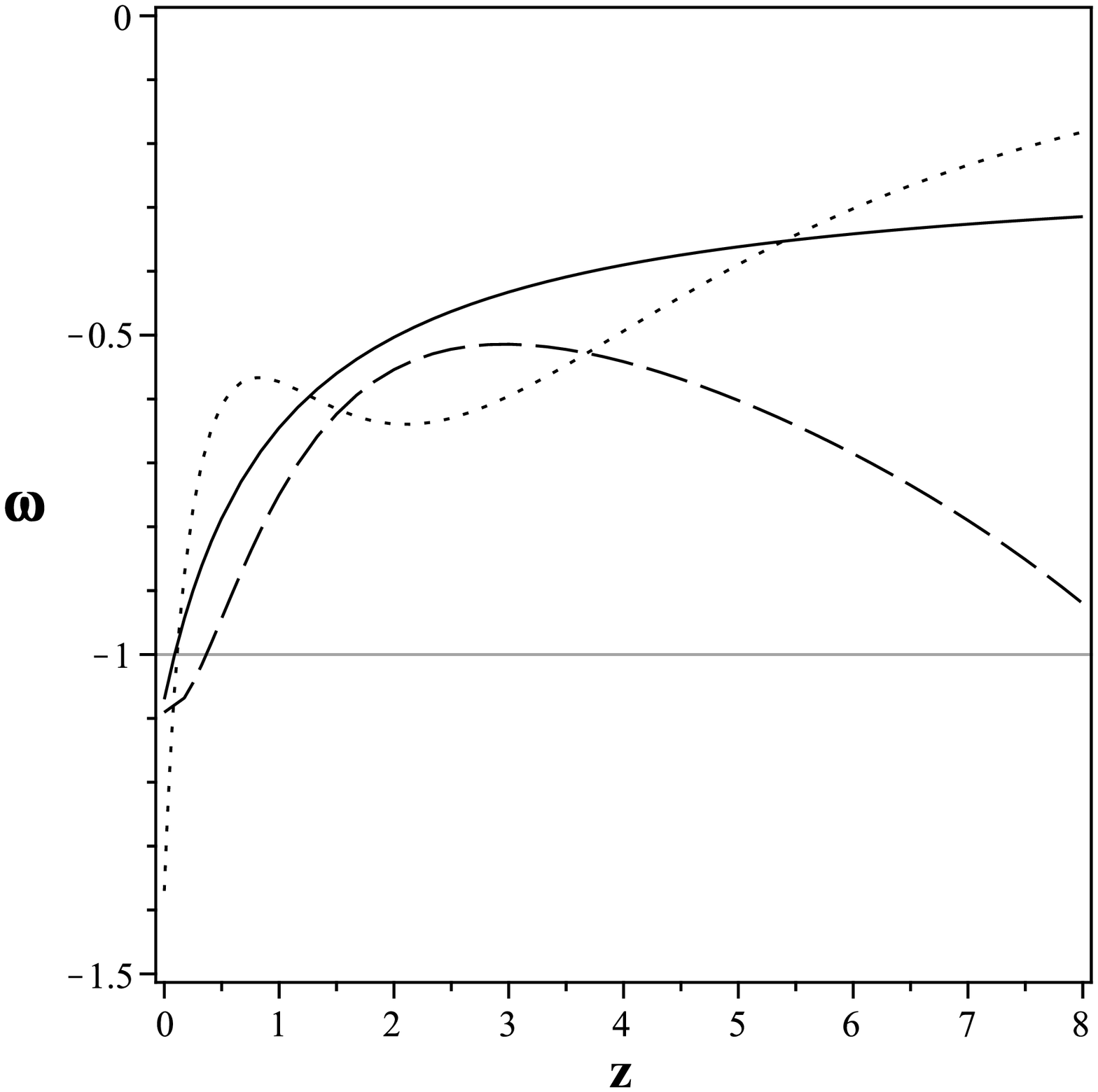,scale=.3}\epsfig{file=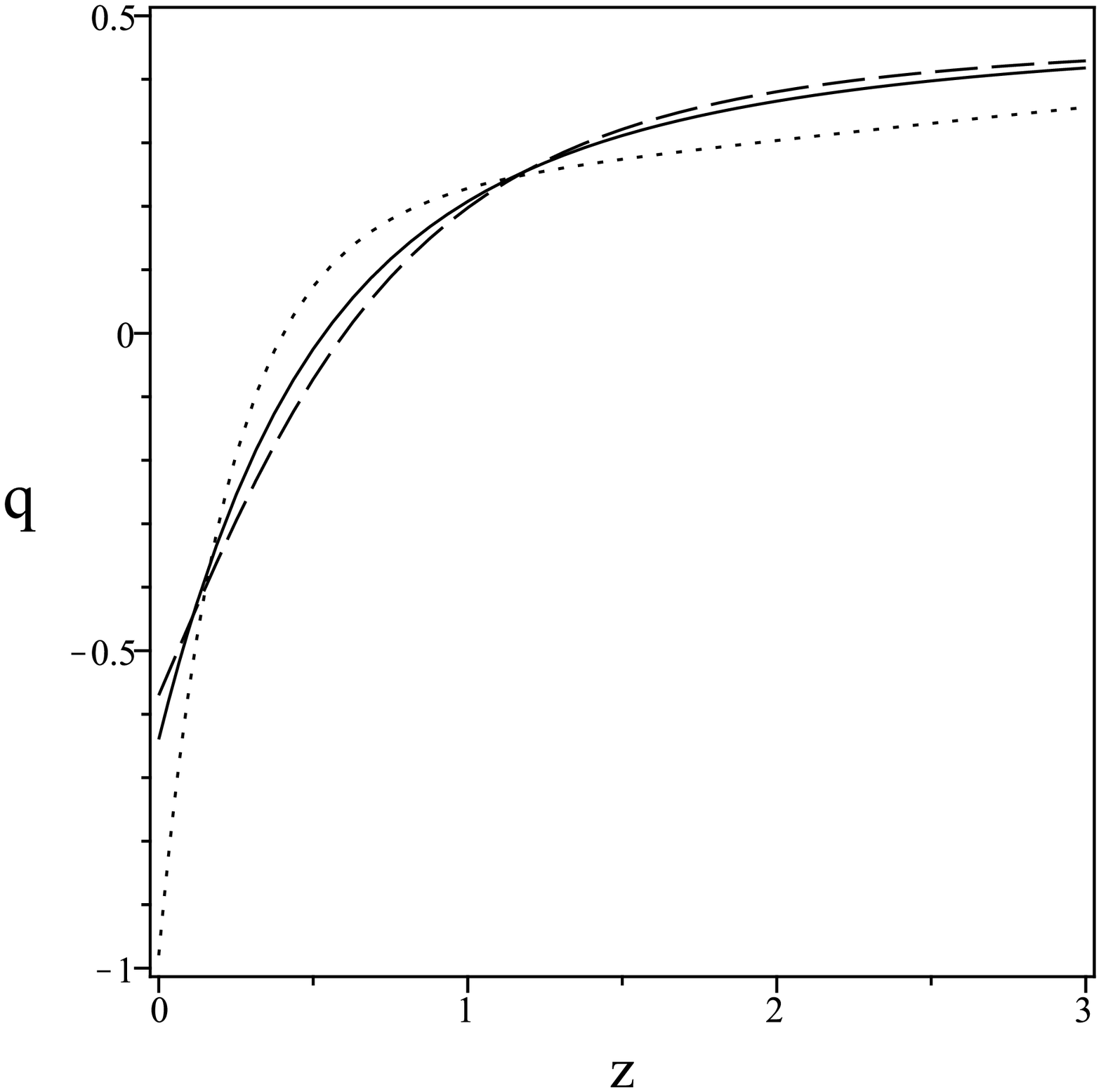,scale=.3}}
\vspace*{8pt} \caption{Graphs of the EoS and deceleration parameter
in respect of redshift $z$. The solid, dot and dash lines represent
parametrization 1, 2 and 3, respectively.}
\end{figure}
\begin{figure}[th]
\centerline{\epsfig{file=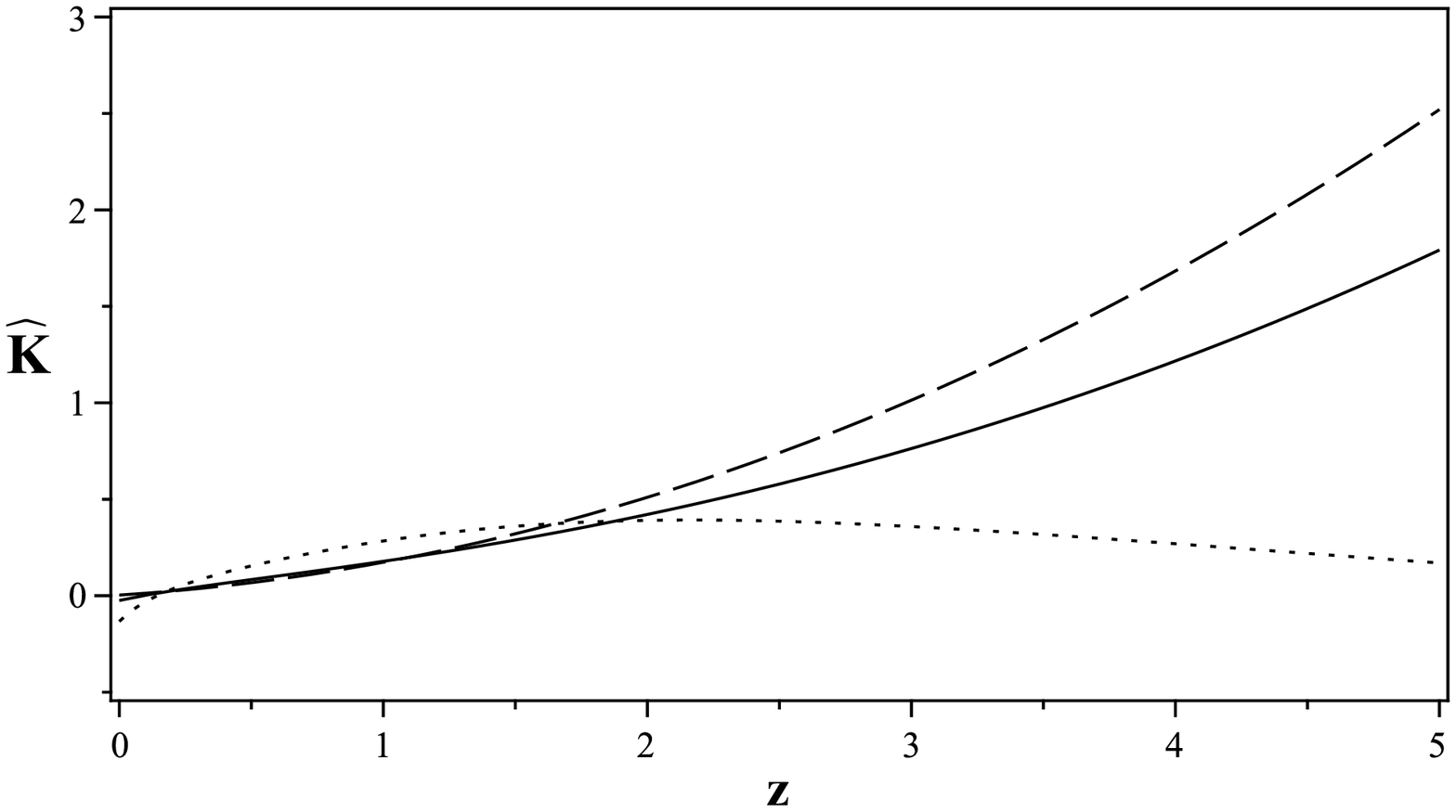,scale=.4}} \vspace*{8pt}
\caption{Graphs of the reconstructed $\hat{K}$ in respect of
redshift $z$. The solid, dot and dash lines represent
parametrization 1, 2 and 3, respectively.}
\end{figure}
\begin{figure}[th]
\centerline{\epsfig{file=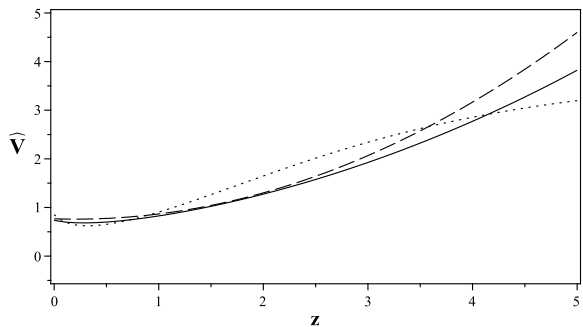,scale=.4}} \vspace*{8pt}
\caption{Graphs of the reconstructed $\hat{V}$ in respect of
redshift $z$. The solid, dot and dash lines represent
parametrization 1, 2 and 3, respectively.}
\end{figure}
From Figs.  4, 5 and 6, we can see parametrization $1$ and $3$ are
same nearly and have slightly different from parametrization $2$.
The EoS for any parametrization  show in Fig. (4) so that they
running cross to $-1$. Acceleration for all of parametrization
shows to tend to the positive value. The $\hat{K}$ and $\hat{V}$
increase for parametrization $1$ and $3$, but in parametrization
$2$ increase (decrease) for the $\hat{V}$ ($\hat{K}$). One can
see that parametrization 1 and 3 satisfy condition
$\hat{K}+\hat{V}>0$ and parametrization 2 satisfy condition
$\hat{K}+\hat{V}=Constant$. In order to by Eq. (\ref{E26-1}), we
have $\omega>-1$ ($\omega=-1$) when parametrization 1 and 3
(parametrization 2). This is mean that
parametrization 2 is better than others parametrization.\\
\begin{figure}[th]
\centerline{\epsfig{file=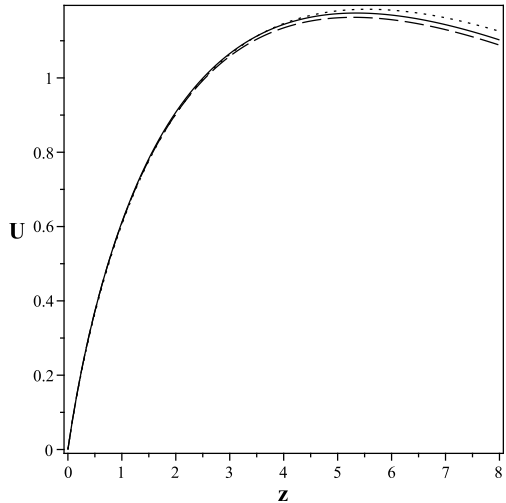,scale=.35}} \vspace*{8pt}
\caption{Graphs of the reconstructed $U_i$ in respect of redshift
$z$. The solid, dot and dash lines represent parametrization 1, 2
and 3, respectively.}
\end{figure}
In Fig. 6, we can see the variation of the vector field against
redshift $z$. One is obviously showed  slightly difference between
all of parametrization.\\\\
\newpage

\section{Conclusion}
In this paper, we have studied the bouncing solution in curved
universe which proposed by the model of a massive vector field,
$U_i$, non-minimally coupled to gravity. For our purpose we have
derived the corresponding energy density, pressure and Friedmann
equation for this model. Also we have obtained the bouncing
condition as Eq.(\ref{E19}). From this condition, and also
essential condition (\ref{E16}), it is clear that if we have
bouncing solutions in open universe, then we have such behaviour
for flat and closed universe as well. After we plot the Hubble
parameter in term of time by figures 2,  for different $k$, we
understood  that our model predict the bouncing behavior for all
cases of $k$. Fro these figures one  can see that the Hubble
parameter $H$ running across zero in any three cases of $k$. In
all cases of $k$, we have $H<0$ to $H>0$ where implies to go from
collapse era to an expanding era, and this result will not change
for the different values $\xi$ in all of $k$. After that in
figures 3 we have shown that during the contracting phase, the
scale factor $a(t)$ is decreasing,
 i.e., $\ddot{a}<0$, and in the expanding phase we have $\ddot{a}>0$, so the point where
$\ddot{a}=0$ is bouncing point, and this figure is consistent
with the results of Fig 2.\\
After that we have investigated an interesting method as the
reconstruction of the non-minimally coupled massive vector field
model with the action (\ref{E1}). Our aim was to see whether the
non-minimal coupling vector field can actually reproduce required
values of observable cosmology, such as evolution of the EoS and
the deceleration parameter in respect to the redshift $z$. We
have reconstructed our model in three different forms of
parametrization for massive vector field. In Fig. 4 we have found
the EoS crossing $-1$ in all of parametrization. The variation of
reconstructed kinetic and potential energy against $z$ have
plotted in Figs. 5 and 6, where the parametrization 2 in addition
is better than two others parametrization because
$\hat{K}+\hat{V}=Constant$. Also we have investigated the
stability of this system and have obtained a condition by the
sound speed in all of curvatures. Finally we note that
reconstructed physical quantities have just executed in flat
universe and one is suggested for open and close universe as
future work.\\\\

\section{Acknowledgment}
 The authors are indebted to the anonymous referee for
his/her comments that improved the paper drastically.

\end{document}